\documentclass[runningheads]{llncs}
\usepackage{graphicx}
\usepackage{hyperref}
\usepackage{cleveref}
\usepackage{booktabs}

\begin{document}
\title{Accurate Detection of Mediastinal Lesions with nnDetection}
%
%
\author{Michael Baumgartner$^{1,4}$, Peter M. Full$^{1,2}$, Klaus H. Maier-Hein$^{1,3}$}
%

\authorrunning{Baumgartner et al.}

%

\institute{%
$^1$Division of Medical Image Computing, German Cancer Research Center, Heidelberg, Germany\\
$^2$Medical Faculty Heidelberg, Heidelberg University, Heidelberg, Germany\\
$^3$Pattern Analysis and Learning Group, Heidelberg University Hospital\\
$^4$Helmholtz Imaging\\
\email{m.baumgartner@dkfz.de}}

\maketitle              
\begin{abstract}
The accurate detection of mediastinal lesions is one of the rarely explored medical object detection problems.
In this work, we applied a modified version of the self-configuring method nnDetection to the Mediastinal Lesion Analysis (MELA) Challenge 2022.
By incorporating automatically generated pseudo masks, training high capacity models with large patch sizes in a multi GPU setup and an adapted augmentation scheme to reduce localization errors caused by rotations, our method achieved an excellent FROC score of $0.9922$ at IoU $0.10$ and $0.9880$ at IoU $0.3$ in our cross-validation experiments. The submitted ensemble ranked third in the competition with a FROC score of $0.9897$ on the MELA challenge leaderboard.

\keywords{Mediastinal Lesions, CT, Object Detection, nnDetection}
\end{abstract}
\section{Introduction}



While many pathologies were already explored for medical object detection by previous studies~\cite{setio2017validation,jin2020deep,ivantsits2022detection,baumgartner2021nndetection}, the detection of lesions in the mediastinum was rarely investigated before. 
The MELA challenge 2022 aims to address this shortcoming by providing a large, publicly available data set and leaderboard to benchmark new advances in the domain.
Our submission to the challenge is based on nnDetection~\cite{baumgartner2021nndetection}, a recently proposed self-configuring object detection method which can be applied to new problems without manual intervention.
It follows nnU-Net's~\cite{isensee2021nnu} design principle to systematise and automate the hyperparameter tuning process by using fixed, rule-based and empirically derived parameters.
Since nnDetection was developed for data sets with pixel-wise annotations and GPUs with 11GB of VRAM, we make minor adjustments to its configuration to fully leverage our available resources and tackle three challenges of this specific data set:


\begin{itemize}
    \item \textbf{Bounding Box Annotations}: Enclosing lesions with bounding boxes, enables the collection of large scale data sets by speeding up the annotation process. However, compared to pixel-wise segmentation, less information is stored in bounding boxes which limits the use of spatial data augmentation transformations and available training signals.

    
    \item \textbf{Large Lesions}: Lesion bounding boxes can only be predicted accurately when the entire lesion is visible in the image patches used for inference. MELA has particularly large lesions, requiring large patch sizes and in turn an immense amount of GPU memory to train appropriate models.
    

    \item \textbf{Accurate Localization}: MELA has an unusually high IoU cutoff at 0.3 (versus the commonly used 0.1 \cite{baumgartner2021nndetection,jaeger2020retina}) requiring particularly precise bounding box localization especially in the case of small lesions.

\end{itemize}

Our proposed solution incorporates automatically generated pseudo masks which were derived from the provided bounding boxes, training high capacity models with large patch sizes in a multi GPU setup and an adapted augmentation scheme to reduce localization errors caused by rotation when training with coarse annotations to tackle the aforementioned challenges.


\section{Methods}
We used nnDetection \cite{baumgartner2021nndetection}, a self-configuring method for volumetric medical object detection, as our development framework.
Some of the fixed parameters were adapted to account for the availability of improved resources and the bounding box annotations while the rule based and empirical parameters were used without modification, see \cref{sec:modifications}.
Since many lesions exceed the patch size of the full resolution models, all of our experiments used the automatically created low resolution data set~\cite{baumgartner2021nndetection}.

\subsubsection{Preprocessing}

Motivated by previous studies by Tang et al. \cite{tang2019uldor} and Zlocha et al. \cite{zlocha2019improving} additional segmentation masks were generated to encode prior information into the training process.
We used the center and size of the bounding boxes to fit ellipsoids which mimic the round shape of lesions.
By leveraging the generated segmentations throughout the whole data preparation and data loading pipeline, we can use spatial transformations from commonly available data augmentation frameworks, such as batchgenerators~\cite{isensee2020batchgenerators}, to artificially diversify the data.
Since multiple CT scans in the data set had a similar spacing, the resampling step was only executed on scans where at least one of the spacings differed more than $5\%$ from the target spacing of $[1.40,1.43,1.43] mm$.



\subsubsection{Training}


Each network is trained for 50 epochs each consisting of 2500 batches.
Gradient updates are performed with Stochastic Gradient Descent (SGD) and a Nesterov momentum term of $0.99$.
During the data loading process, a random offset is applied to each dimension where the object size does not exceed $70\%$ of the patch size while making sure that the object remains fully contained inside the patch.
The maximum offset is limited to $70\%$ of the difference between the object size and patch size.
For object dimensions which exceed that threshold, a random center point is selected inside the object boundaries.

\subsubsection{Network Topology}

The network topology follows the original nnDetection implementation and consists of convolutions with non-linear activation and normalisation layers.
The ReLU activations were replaced with Leaky ReLU activations to avoid dead neurons during the training process.
We kept Retina U-Net~\cite{jaeger2020retina} as our blueprint architecture for all experiments which results in three training branches: anchor classification trained with the Binary Cross Entropy loss, anchor regression trained with a weighted L1 loss and a semantic segmentation branch which is trained with the Cross Entropy and Dice loss.
The total loss composition can be summarised as follows:

\begin{equation}
  L_{total} = L_{BCE} + 2 \cdot L_{L1} + L_{CE\_seg} + L_{Dice\_seg}
\end{equation}

\begin{figure}[t]
    \centering
    \includegraphics[width=\textwidth]{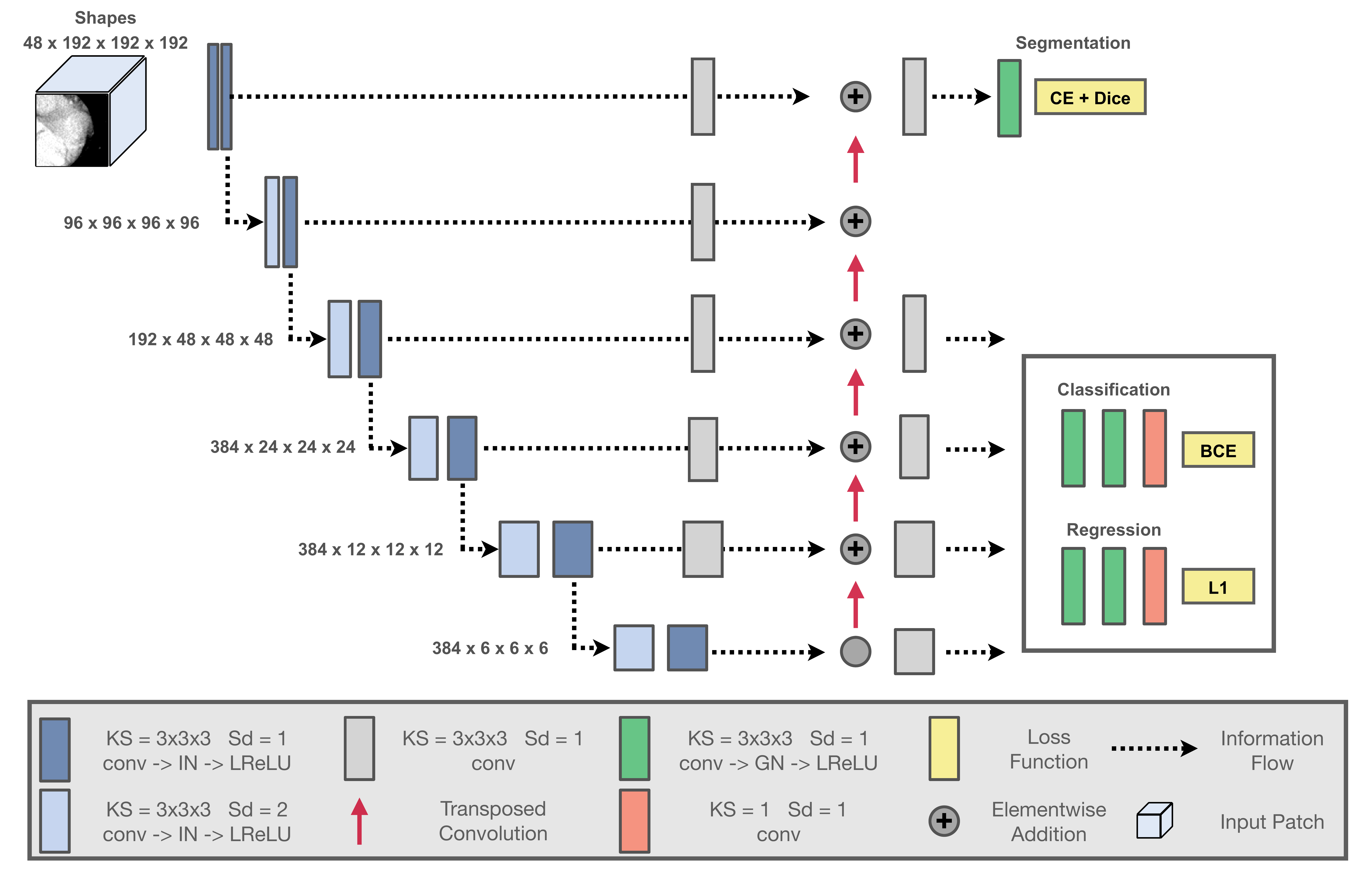}
    \caption{Depicts Retina U-Net, the blueprint architecture of nnDetection which is automatically configured based on the data set properties. The patch size was manually adjusted to account for the large objects present in the MELA data set. The encoder architecture is based on stacked convolutions while the decoder is based on the Feature Pyramid Network~\cite{lin2017feature}. Instead of bi-linear interpolation, transposed convolutions are used to learn the up-sampling. Abbreviations: conv = Convolution, KS = Kernel Size, Sd = Stride, IN = Instance Normalisation, GN = Group Normalisation~\cite{wu2018group}, LReLU = Leaky Rectified Linear Unit}
    \label{fig:architecture}
\end{figure}

To increase the model capacity, the number of channels were increased by $50\%$ and the maximum number of channels was set to $384$.
A detailed overview of the resulting architecture is shown in~\cref{fig:architecture}.

\subsection{Challenge Specific Modifications}
\label{sec:modifications}
In order to fully leverage our available resources and incorporate task specific aspects of the MELA challenge, we performed minor manual changes to our baseline.

\subsubsection{Large Patch Size (LP)}
While the automatically generated low resolution data set substantially increases the contextual information for most lesions, some predictions did not exceed the required IoU threshold for very large ones.
This was caused by errors when combining predictions of the same object between neighboring patches during inference.
To reduce these stitching artifacts, we increased the patch size from $[160, 128, 128]$ to $[192, 192, 192]$.
Due to the cubic increase in the required VRAM, these models were trained on two Nvidia A100 (40 GB) with batch size $4$ per GPU.


\subsubsection{Reduced Rotation during Augmentation}

\begin{table}[htp]
    \centering
    \begin{tabular}{ c | c | c }
     
     Augmentation & Baseline (Aug A) & Reduced Rotation (Aug B) \\
     \hline
     Elastic Deformation & $\times$  & $\times$ \\
     Rotation (m in degrees) & p=0.3 m=[-30, 30] & p=0.1 m=[-10, 10] \\
     Scaling & p=0.2 m=[0.7, 1.4] & p=0.3 m=[0.65, 1.6] \\
     Rotation 90 & $\times$ & p=0.5 \\
     Transpose Axes  & $\times$ & p=0.5 \\
     Gaussian Noise & p=0.1 & p=0.1 \\
     Gaussian Blur & p=0.2 & p=0.2 \\
     Median Filter & $\times$ & p=0.2 \\
     Multiplicative Brightness & p=0.15 & $\times$ \\
     Brightness Gradient & $\times$ & p=0.3 \\
     Contrast & p=0.15 & p=0.2 \\
     Siumlate Low Resolution & $\times$ & p=0.15 \\
     Gamma & p=0.3 & p=0.1 \\
     Inverse Gamma & p=0.1 & p=0.1 \\
     Local Gamma & $\times$ & p=0.3 \\
     Sharpening & $\times$ & p=0.2 \\
     Mirror (per axes) & p=0.5 & p=0.5 \\

    \end{tabular}
    \caption{Shows two different augmentation schemes used throughout our experiments. The probability that an augmentation is applied is denoted by $p$ while the magnitude is denoted by $m$.}
    \label{tab:augmentation}
\end{table}

Spatial data augmentations are a central part of modern deep learning pipelines to reduce overfitting and improve generalisation of the developed methods.
By applying rotations and scaling to the images, the diversity in the data set is artificially increased by training on augmented versions of the same patch.
Rotations which are not a multiple of 90 degrees can be troublesome when coarse annotations are used during training since the axis aligned bounding boxes can not be derived without introducing localization errors.
To account for this fact, a second augmentation pipeline with reduced rotation transformations was used to train a second model.
By adding additional intensity based augmentations, as well as transposing and rotation of 90 degrees we tried to maintain a diverse set of augmentations.
The full list of the augmentations, probabilities and magnitudes is shown in~\cref{tab:augmentation}.




\section{Experiments and Results}
The MELA data set provides 770 training and 110 validation CT scans.
To reduce overfitting during the development phase, all scans were merged into one pool and 5-fold cross-validation was used to train and evaluate the models.
Performance is measured by the FROC score with sensitivities computed at $0.125, 0.25, 0.5, 1, 2, 4, 8$ False Positives per Image at an Intersection over Union (IoU) threshold of $0.3$.
Additionally, we report results at an IoU threshold of $0.1$ for our cross-validation experiments.
During the final phase, the test data was released to the participants and up to five submission per day were allowed.
Our final submission was based on the ensemble of the two training strategies with large patch sizes and different data augmentation schemes.
An overview of all the results can be found in \cref{tab:results}.

    
     

\begin{table}[htp]
    \centering
    \begin{tabular}{lcccccc}
    \toprule
     Model           & Patch Size    &  Batch Size    & Augmentation &  \multicolumn{2}{c}{5 Fold CV} & Leaderboard \\\cmidrule(lr){5-6}
                     &               &                &              & IoU 0.1     & IoU 0.3 & \\
    \midrule
      Baseline       & [160,128,128] & 6              & A            & 0.9844       & 0.9808              & 0.9824 \\ 
      LP           & [192,192,192] & 8              & A            & 0.9858       & 0.9809              & 0.9851 \\ 
      Aug B, LP   & [192,192,192] & 8              & B            & \textbf{0.9927}       & 0.9846              & 0.9851 \\ 
      LP Ensemble    & [192,192,192] & 8              & A+B          & 0.9922 & \textbf{0.9880} & \textbf{0.9897} \\ 
    \bottomrule
    \end{tabular}
    \caption{Results of different training strategies for our 5-fold cross-validation experiments and leaderboard submissions. Differences in patch size, effective batch size and data augmentation are listed in separate columns. The LP ensemble denotes the ensemble of the two training strategies with large patch sizes.}
    \label{tab:results}
\end{table}

\begin{figure}[t]
    \centering
    \includegraphics[width=\textwidth]{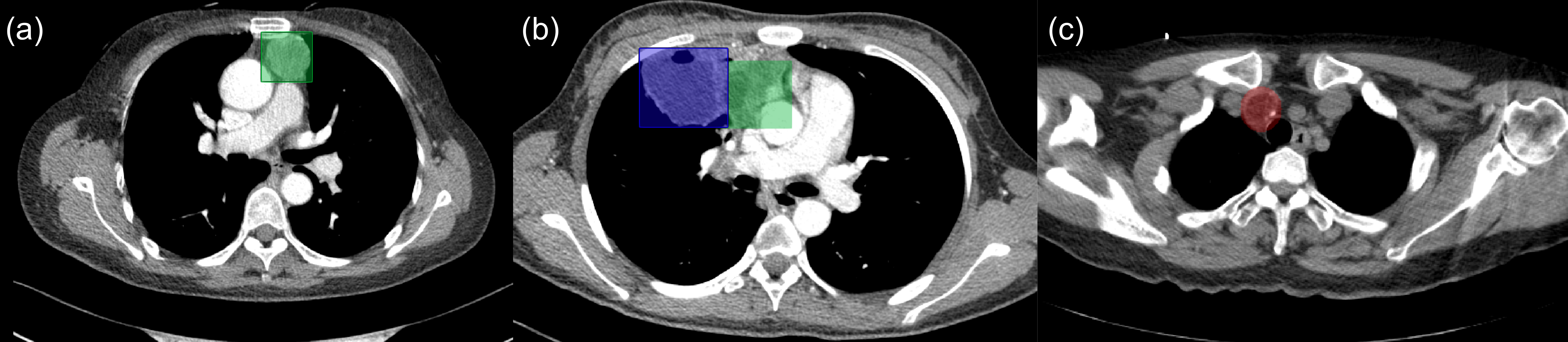} 
    \caption{Qualitative results of the final ensemble model. (a) shows an accurately detected lesion (green) (b) contains one correctly predicted
    lesion (green) and one false positive prediction (blue) of a lung nodule (c) shows a missed lesion (red).}
    \label{fig:qual_results}
\end{figure}

All models achieved an excellent FROC score above $0.98$ while a performance gap between the lower and higher IoU thresholds persisted throughout our cross-validation experiments.
Using a large patch size for additional contextual information showed slight improvements during the cross-validation and the leaderboard.
Introducing the second augmentation scheme improved the results for both the high and low IoU threshold in the cross-validation setting but did not yield improved results on the leaderboard.
Finally, the ensemble of the two models with large patch sizes showed the best performance for an IoU threshold of $0.3$ in our cross-validation and the leaderboard.
The gap between the cross-validation and leaderboard results remains small for all of our models, showing good generalisation performance to unseen data.
Some qualitative results for True Positive, False Positives and False Negatives from our cross-validation experiments are shown in~\cref{fig:qual_results}.

Based on our qualitative analysis the detection of smaller lesions remains more difficult than the detection of large lesions.
This might be partially caused by the downsampled data set to improve the detection performance of large lesions.
Surprisingly, the model was even able to correctly predict a lung lesion which had a similar appearance as the neighboring mediastinal lesion.


\section{Discussion}
We presented our solution to the MELA 2022 challenge which was based on nnDetection and showed good results with little modifications to its pipeline.
By using automatically generated pseudo masks, training with large patch sizes in a multi GPU setup and ensembling two models with different augmentation schemes our method achieved an excellent FROC score of $0.9922$ at IoU $0.10$ and $0.9880$ at IoU $0.3$ in our cross-validation experiments and a FROC score of $0.9897$ on the leaderboard.
Our submission ranked third in the MELA challenge 2022.
While the detection of mediastinal lesions in this data set can already be solved with very high performance, the simultaneous detection of small and large objects remains a difficult challenge.
The integration of prior knowledge into models to improve their regression performance and enhancing the post-processing procedures to improve the locatization performance of small lesions might be promising directions for future research.



\section*{Acknowledgements}
Part of this work was funded by Helmholtz Imaging, a platform of the Helmholtz
Incubator on Information and Data Science.

%
%
%
%

\bibliographystyle{splncs04}
{\bibliography{references}}


\end{document}